\pdfoutput=1
\pdfoutput=1
\pdfoutput=1
\pdfoutput=1
\pdfoutput=1

\documentclass[conference,a4paper]{IEEEtran}

\IEEEoverridecommandlockouts

\usepackage{ifpdf}

\usepackage{cite}
\usepackage{amsmath,amssymb,amsfonts}
\usepackage{algorithmic}
\usepackage{graphicx}
\usepackage{textcomp}
\usepackage{xcolor}
\def\BibTeX{{\rm B\kern-.05em{\sc i\kern-.025em b}\kern-.08em
    T\kern-.1667em\lower.7ex\hbox{E}\kern-.125emX}}

\begin{document}

\title{{Detection of False Data Injection Attacks Using the Autoencoder Approach}
\thanks{This work is supported by the Chinese Scholarship Council.}
\author{\IEEEauthorblockN{Chenguang~Wang, Simon~Tindemans, Kaikai~Pan, Peter~Palensky\\}
\IEEEauthorblockA{\textit{Department of Electrical Sustainable Engineering} \\
\textit{Delft University of Technology}\\
Delft, The Netherlands \\
\{c.wang-8,  s.h.tindemans, k.pan, p.palensky\}@tudelft.nl}}}

\IEEEoverridecommandlockouts

\IEEEpubid{\begin{minipage}{\textwidth}\copyright 2020 IEEE.  Personal use of this material is permitted.  Permission from IEEE must be obtained for all other uses, in any current or future media, including reprinting/republishing this material for advertising or promotional purposes, creating new collective works, for resale or redistribution to servers or lists, or reuse of any copyrighted component of this work in other works.
\end{minipage}}

\maketitle

\IEEEpubidadjcol

\begin{abstract}
State estimation is of considerable significance for the power system operation and control. However, well-designed false data injection attacks can utilize blind spots in conventional residual-based bad data detection methods to manipulate measurements in a coordinated manner and thus affect the secure operation and economic dispatch of grids. In this paper, we propose a detection approach based on an autoencoder neural network. By training the network on the dependencies intrinsic in ‘normal’ operation data, it effectively overcomes the challenge of unbalanced training data that is inherent in power system attack detection.  To evaluate the detection performance of the proposed mechanism, we conduct a series of experiments on the IEEE 118-bus power system. The experiments demonstrate that the proposed autoencoder detector displays robust detection performance under a variety of attack scenarios. 
\end{abstract}


\section{Introduction}\label{sec:intro}
The power system is increasingly equipped with sensors and communication infrastructures. This enables smarter grid operations, but also makes possible novel cyber attack scenarios that  manipulate power system measurements instead of directly disrupting ICT infrastructure or stealing valuable data. Although the typical bad data detection (BDD) within state estimation (SE) can detect erroneous measurements and some “basic” attacks, well-designed attacks can remain stealthy and bypass the BDD, such as the stealthy false data injection attacks (FDIAs) \cite{liu2011false}. These stealthy measurements manipulation attacks severely threaten both the economic dispatching and security control of the power system \cite{liu2015analyzing, jia2013impact}. 

Several techniques have been proposed to deal with stealthy FDIAs. In \cite{manandhar2014detection}, the authors have proposed a Kalman filter estimator together with a chi-square detector. Other statistical methods, such as sequential detection using Cumulative Sum (CUSUM)-type algorithms were designed in \cite{Li2015}. The recent work \cite{Zhao2018} has proposed a detector utilizing the statistical consistency of measurements, presuming that the system is observable by a minimal set of secure phasor measurement units. These methods, however, can be limited by the prior assumption that measurements fit specific distributions, or by restrictions on the number of manipulated measurements \cite{Pan2019a}.

\IEEEpubidadjcol
\IEEEpubidadjcol

Moreover, it is increasingly recognised that the distribution of normal power system states is not easily characterised using standard parametric distributions \cite{sun2016evaluating}. The need to operate in a complex stochastic environment has led to the deployment of data-driven methods. For example, distance-based algorithms like $k$ nearest neighbour (k-NN) were used to cluster normal and corrupted measurement states \cite{tian2014anomaly}. Nevertheless, the very high dimensionality of measurements (from the physical, cyber and market domains) results in data sparsity, where manipulated measurements may be masked by the noise of multiple irrelevant dimensions. This can make detection using a high-dimensional distance-based algorithm computationally inefficient or even invalid \cite{aggarwal2015outlier}.

Alternative data-driven approaches to FDIA detection have been proposed in the form of support-vector machine (SVM)-based classifiers \cite{he2017real} and deep neural network-based classifiers \cite{james2018online}. Both are supervised machine learning algorithms that classify measurements into normal and manipulated data on the basis of labeled training data. However, due to the infrequent occurrence (or more likely: absence) of FDIAs in historical data, the training data set is highly unbalanced, so that it must be augmented by simulated training data. Moreover, in this way, the detector only learns to detect known attacks, which is a significant weakness in a fast-evolving field with resourceful and potentially well-equipped attackers. 

This paper bridges the identified gap by proposing a detection approach based on an autoencoder neural network. The main contributions of this paper are listed below:
\begin{itemize}
	\item[1)] We propose an  autoencoder-based detection approach for FDIAs. It learns to identify anomalous system states (and therefore possible attacks) using only SCADA-type power flow measurements for a large range of normal operating conditions. Therefore it is well-suited to the inherent data imbalance in FDIA detection applications. 
	\item[2)] We define a case study on the IEEE 118-bus system, including a model to generate `normal' data. We formulate two FDIA scenarios 
	by considering comprehensive factors of the adversaries’ purpose, capacity, and knowledge and utilize indicators to evaluate the FDIA detection performance of our proposed mechanism. The experimental results demonstrate the mechanism has a satisfactory detection accuracy.
\end{itemize}

\section{State Estimation and Data Attacks} \label{sec:sebdd}

In this section, we briefly review the state estimation and bad data detection technique and formulate the FDIA problem.

\subsection{State estimation} \label{subsec:se}

The power system we consider has $n_{b}$ buses and $n_{t}$ transmission lines. The vector ${\theta} = [ \theta_{1}, \, \theta_{2}, \, \ldots, \, \theta_{n_b} ]^{T} $ represents $n_{b}$ phase angles, excluding the angle of the reference bus. In this paper, a DC power flow model is assumed, in which the reactive power is neglected and bus voltages are assumed to be 1 (p.u.). The vector ${P^{{I}}} \in \mathbb{R}^{n_{b}}$ of active power injections is related to the angle vector ${\theta}$,
\begin{align}\label{eq:pipf}
{P}^{{I}} = A P^{F} = AR^{-1}A^{T}\theta,
\end{align}
where $P^{F}  \in \mathbb{R}^{n_{t}}$ is the branch active power flow vector, $R \in \mathbb{R}^{n_{t} \times n_{t}}$is a diagonal matrix of transmission line reactances and $A \in \mathbb{R}^{n_{b} \times n_{t}}$ is the branch-to-node incidence matrix \cite{gonzalez2014powerfactory}. In the following, we shall use the power injection vector $P^{I}$ as the system state $ x \in \mathbb{R}^{n_{b}}$. It is functionally equivalent to the more commonly used phase angle vector $ \theta$, but it allows for more elegant generation and detection of FDIAs. 

We consider a system where the active power injections and line flows are measured with some error. Thus the system model $H \in \mathbb{R}^{(n_{b}+n_{t}) \times n_{b}} $ for measurement and state can be written by 
\begin{align}\label{eq:z}
z = \left[\begin{matrix} I \\ H^{F} \end{matrix}\right]x + e = Hx + e,
\end{align}
where the measurement noise vector $e \sim \mathcal{N}(0, \, D)$ denotes $m$ independent zero-mean Gaussian variables with the covariance matrix $D = \mbox{diag}(\delta_{1}^{2}, \, \ldots, \, \delta_{m}^{2}) $ and the measurement vector $z \in \mathbb{R}^{m}$ indicates measured power injection and line power flow with noise. Identity matrix $I \in \mathbb{R}^{n_{b} \times n_{b}}$ and distribution factor matrix $H^{F} \in \mathbb{R}^{n_{t}\times n_{b}}$ are parts in $ H $ corresponding to the power injection and line power flow, respectively. According to \eqref{eq:pipf}, the distribution factor matrix can be described as $H^{F} = R^{-1}A^{T}(AR^{-1}A^{T})^{-1}$. Given the observation of the measurements $z$, the state estimate $\hat{x}$ is solved by the weighted least squares (WLS) approach \cite{sandberg2010security} as

\begin{align}\label{eq:hatx}
\hat{x} = (H^{T}D^{-1}H)^{-1}H^{T}D^{-1}z := Kz.
\end{align}

\subsection{Bad data detection and stealth FDIAs} \label{subsec:bddfdia}

The vector $ \hat{x}$ is then utilized to estimate the power injection and line power flow measurements by $\hat{z} = H\hat{x} $. In bad data detection, a residual is defined to describe the difference between the actual and the estimated measurements, namely $ r_{o} = z - \hat{z}$. This naturally gives rise to a BDD scheme that identifies bad data by comparing the 2-norm of $r_{o}$ with a certain threshold $\tau$, i.e. an alarm is triggered if $\| r_{o} \|_{2} > \tau$.

We denote $a \in \mathbb{R}^{m}$ as the non-zero false data vector injected into measurement vector $z$. The manipulated measurement vector can be described as $z_{a} = z + a$. Here the vector $c$ is defined as the deviation of the estimated state before and after the attack. The corrupted system state can be denoted as $\hat{x}_{a} = \hat{x} + c$. According to \eqref{eq:hatx}, the falsified state estimate $\hat{x}_{a}$ can be written by
\begin{align}\label{eq:hatxa}
\hat{x}_{a} & = (H^{T}D^{-1}H)^{-1}H^{T}D^{-1}z_{a} \nonumber\\
& = (H^{T}D^{-1}H)^{-1}H^{T}D^{-1} (z + a) \\
& = \hat{x} + {c}, \nonumber
\end{align}
and the corresponding $r_{a}$ after the attack can be expressed as 
\begin{align}\label{eq:ra}
\begin{array}{ll}
{r}_{a} & = z_{a} - H\hat{x}_{a} = z + a - H(\hat{x} + c) \\
& = r_{o} + (a -Hc).
\end{array}
\end{align}
If $a=Hc$, then the manipulated residual $r_{a}$ equals the original residual $r_{o}$. Thus the attacker manipulates the measurements with the residual unchanged and keeps stealthy with respect to this physics-based BDD scheme. This remains true if $a \neq Hc$, as long as $\| r_{a} \|_{2} \leq \tau$ is still satisfied.

For our FDIA detection study, we consider one attack scenario from the perspective of an adversary that manipulates load patterns \cite{jia2013impact}, for example in order to hide excessive power consumption or to reduce apparent power consumption for economic motives. The adversary needs to corrupt the power generation and power flow accordingly to avoid detection by BDD. The attack scenario will be detailed in section IV.

\section{FDIA Detection Mechanism} \label{sec:detect}

In this section, we propose an FDIA detection mechanism based on the autoencoder algorithm. We first analyze the specific characteristics and advantages of the method for identifying FDIAs in the context of the power system. Then, we explain the attack detection principle of the autoencoder-based mechanism in detail. Finally, we describe the complete training and detection process of our proposed mechanism.

\subsection{Autoencoder-based attack detector} \label{subsec:ae_detect}

FDIA detection is essentially a classification problem with the objective of distinguishing false data from data that is considered ‘normal’. What the SVM-based \cite{he2017real} and deep neural network-based classifiers \cite{james2018online} have in common is to treat FDIA detection as a supervised learning task. However, supervised learning requires a training data set with representative examples of normal system operation and attacks. Such data sets are in short supply, because of the rarity of attacks, unwillingness to share data, and evolving attacks. As a result, it is difficult to learn a satisfactory discriminator of ‘normal’ and ‘attack’ scenarios on this basis \cite{duan2016new}. 

Instead, we propose to approach FDIA detection as a one-class classification problem, where the detector is trained on examples of only ‘normal’ operation data. Observations with features that deviate substantially from those in the training data will be considered anomalies, in this case as ‘potential attacks’. There are two main advantages to this approach. First, the autoencoder-based mechanism avoids the need to gather or generate attack data to create balanced data sets for training the classifiers. Second, by focusing on what is normal only, the proposed mechanism is naturally prepared for unknown attack patterns.

\begin{figure}[t!p]
	\centering
	\includegraphics[scale=0.661]{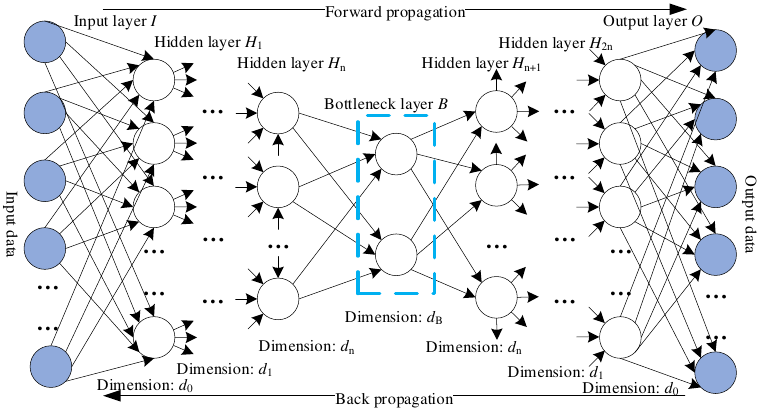}
	\caption{The schematic of the Autoencoder.}
	\label{fig:autoencoder}
\end{figure}

Autoencoders learn the most important features of the training data (i.e. normal power system measurements) by sending the measurements through an information bottleneck while attempting to reconstruct the training data with minimal error \cite{sakurada2014anomaly}. The structure of the autoencoder algorithm is depicted in Fig.~\ref{fig:autoencoder}. The dimension reduction process of mapping the $d_{0}$-dimensional input data to the code in the bottleneck layer $B$ through hidden layers $H_{1}$ to $H_{n}$ is named the \emph{encoder}. Afterwards, the \emph{decoder} decompresses the code to $ d_{0}$-dimensional output data. Weight matrices $W$ and bias vectors $b$ are utilized in the encoding and decoding process as 
\begin{subequations}
	\begin{align}\label{eq:y}
	Y = \sigma(W_{n}^{e}(\ldots \sigma(W_{0}^{e}Z + b_{0}^{e})\ldots) + b_{n}^{e}) \, , 
	\end{align}
	\begin{align}\label{eq:hatz}
	\hat{Z} = \sigma(W_{n}^{d}(\ldots \sigma(W_{0}^{d}Y + b_{0}^{d})\ldots) + b_{n}^{d}) \, , 
	\end{align}
\end{subequations}
where $W_{n}^{e}$ and $W_{n}^{d}$ denote weight matrices for encoding and decoding process respectively, $b_{n}^{e}$ and $b_{n}^{d}$ are bias vectors, and $\sigma$ represents a nonlinear element-wise activation function. $Z$ refers to the input data vector, $Y$ is the data in the bottleneck layer and vector $\hat{Z}$ stands for the output data. 

\subsection{Training and detection process} \label{subsec:train_test}

The residual associated with a training observation $Z_{j}$ is given by $ {r}_{j} = {Z}_{j} - \hat{Z}_{j}$.
The reconstruction error $R_j$ is expressed as the ratio of the length of $r_j$ to the input data dimension $d_0$ and the objective of the training process is to minimize the mean value of the sum of all reconstruction errors ${R}_{j}$  as
\begin{align}\label{opt:min_sum}
&\begin{array}{ccl}
& \min\limits_{W, \, b} & \Big\{ J := \frac{1}{S} \sum\limits_{1}^{S} \, (\lVert {r}_{j}\rVert^{2}/d_0) \Big\} \, ,  \\
\end{array}
\end{align}
where $S$ denotes the total number of the observations used for training. By training the autoencoder on training data that is considered normal, it learns to efficiently encode the features of this data in the bottleneck layer $B$. Data that deviates from the training data in a structural way is therefore highly likely to have a larger reconstruction error. 

\begin{figure}[t!p]
	\centering
	\includegraphics[scale=0.85]{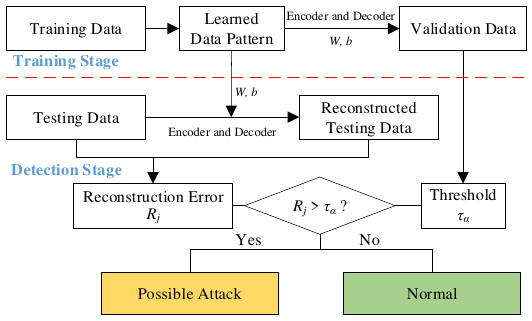}
	\caption{The proposed training and FDIA detection mechanism.}
	\label{fig:flowchart}
\end{figure}

The training and FDIA detection process of the proposed mechanism is depicted in Fig.~\ref{fig:flowchart}. In the training stage, the algorithm iteratively updates the value of weight matrices $W$ and bias vectors $b$ until the function $J$ converges. At the end of the training process, the reconstruction errors ${R}_j$ for the \emph{validation} set are sorted in ascending order. A threshold $\tau_{\alpha}$ equals to the $\alpha^{th}$ percentile is then chosen, for example at the value where an ‘inflection point’ occurs in the error distribution. A possible FDIA is detected when, for a measurement $Z_j$ in the \emph{test} set, the reconstruction error $ {R}_j$ exceeds the threshold $\tau_{\alpha}$.

\section{Case Study} \label{sec:casestudy}

In this section, we evaluate the detection performance of the proposed mechanism using a case study on the IEEE 118-bus system. First, we describe the process of modelling normal operating conditions and explain how to create anomalous attack scenarios. Then, we describe and analyse the load-targeted attack scenario. For this scenarios, we will first quantify the detection performance of our proposed detection mechanism. Specifically, the detection probability, false positive rate, false negative rate are tested. Next, the detection performance of our detector will be compared with a conventional BDD detector. To do so, we introduce “knowledge limited” attacks that both detectors can potentially detect. Notably, the “knowledge-limited” attacks are more of interest in reality as the attacker may have an inaccurate (e.g. out-dated or estimated) system model.

\subsection{Modeling normal operating conditions} \label{subsec:norm_operate}

With the long-term secure and stable operation, the power system has a large number of normal operating conditions which involve a significant volume of loads, power generations and power flows data set. Trained by these data, the proposed mechanism will acquire the data pattern which represents the model of normal system operating conditions.

In the IEEE 118-bus system, electricity is supplied by $M=54$ generators,  transmitted via $Q=186$ branches and ultimately consumed by $N=99$ loads. We generate `normal' (i.e. physically feasible and economically reasonable) power system states and corresponding measurements by using optimal power flow solutions.

Second order polynomial cost functions were assumed for generators, i.e., $f(P_{g}^{G}) = C_{g,2}(P_{g}^{G})^{2} + C_{g,1} P_{g}^{G}$. Hence the economic dispatch $P^{G^{*}}$ is solved with the objective to minimize the total generation cost. The solutions are implicitly parameterized by the nodal load $P^L_l$ and generation cost parameter as 
\begin{align}\label{opt:opf}
P^{G^{*}} & =  \mathrm{arg}\min\limits_{P^{G}} \quad \sum\limits_{g = 1}^{M} C_{g,2}(P_{g}^{G})^2 +C_{g,1} P_{g}^{G}   \\
& \mbox{s.t.} \quad \sum\limits_{g = 1}^{M} P_{g}^{G} - \sum\limits_{l = 1}^{N} P_{l}^{L} = 0, \nonumber
\end{align}
where the injection $P^I = P^{I} (P^G, P^L)$ is determined by the mapping of load $P^L$ and generation $P^G$ onto the nodes. 

Normal operating conditions are generated using a data set that contains a total of 43,717 historical hourly loads from 32 European countries between 2013 and 2017 \cite{Muehlenpfordt2019}. These time series were used to generate a 99 load point time series as follows. The national load time series are first divided by 1000, to obtain reasonable magnitudes for individual buses. Then each load point is assigned a random linear combination of the 32 sources by sampling from the Dirichlet distribution with vector valued parameter $(1,\ldots,1)^T$, which generates a uniform distribution on the $31$-simplex. Additionally, a normally distributed variation with a standard deviation of $\pm$5$\%$ of the measured value is added to each measurement.

An additional source of randomness was created by randomly sampling the generating cost coefficients of the 54 generators as follows. Coefficients $C_{g,2}$ were sampled uniformly in the range $[0.085, \, 0.1225]$ $\$/\mathrm{MWh}^2$ and $C_{g,1}$ uniformly in the range $[1,5]$ $\$/\mathrm{MWh}$. These values span the range of generators included in the IEEE 9-bus system supplied with Matpower \cite{zimmerman1997matpower}.

The procedure above was used to generate snapshot injections $P^I = P^{I} (P^{G^*}, P^L)$, which were converted into line flow measurements using $P^F = H^F P^I$. In this investigation, line transmission limits and generator capacities are not enforced, as the focus of this work is on the recognition of load, generation and power flow patterns. This results in a 339-dimensional measurement vector for training,  containing 99, 54 and 186-dimensional data of loads, power generations and line power flows, respectively. Independent measurement noise $e$ is added using a truncated Gaussian distribution with zero mean, standard deviation of 0.33\% and an absolute value less than 1$\%$ of the original value \cite{he2013online}. The generated data set $T \in \mathbb{R}^{43717 \times 339}$ was divided into a training set $T_{r} \in \mathbb{R}^{26197 \times 339}$, a validation set $T_v \in \mathbb{R}^{8760\times 339}$ and testing set $T_e \in \mathbb{R}^{8760 \times 339}$. 

 In this paper, the autoencoder network contains 4 hidden layers in the encoder with dimensions of 339, 256, 128 and 64, respectively. The bottleneck layer has 32 nodes, and the decoder maps the 32-dimensional data to a 339-dimensional output through 3 hidden layers with the same dimensions as the encoder. In this paper, we used the sigmoid activation function between the second and penultimate hidden layer and the Adam Optimizer \cite{kingma2014adam} to iteratively optimize the value of weight matrices $W$ and bias vectors $b$. The batch size and learning rate for training was 256 and $10^{-5}$ respectively and 2000 training epochs were used. Training and testing of the autoencoder was conducted using \texttt{tensorflow} on the Google Colab environment using the GPU option. An initial performance analysis of hyperparameter settings for the autoencoder-based FDIA detector is available in  \cite{Chenguang2020-2}.

\subsection{Creating attack scenarios} \label{subsec:attacks}

We develop feasible FDIAs from the perspective of the adversaries by adding an offset to the normal operating conditions created in the previous section. 
To gain economic profit, attackers inject false data into the grid by using the acquired knowledge of the targeted power system. In the context of this paper, this knowledge is represented by the incidence matrix $A$ (topology) and the reactance matrix $R$ of the transmission lines. Moreover, we assume that the capacity of an attacker is limited by the attackable measurement set \cite{liu2011false} and the maximum number of the measurements that the attacker can corrupt simultaneously.

In the following, we quantify the factors described above. According to the attack capacity, the adversary selects a set of attacked loads  $\mathcal{L}^{A} \subseteq \mathcal{L}$. The attacker then determines the change rate $\beta_{l}$ of each selected load and calculates the total load change $\sum_{l\in \mathcal{L}^{A}} \beta_{l} P_{l}^{L}$, in which $\beta_{l}P_{l}^{L}$ equals the change $\Delta P_{l}^{L}$ of each load. Similarly,  the attack selects a set of attacked generators $\mathcal{G}^{A} \subseteq \mathcal{G}$. Next, the attack determines ratios of the power generating’s change amount $\lambda_{1}: \lambda_{2}: \ldots : \lambda_{|\mathcal{G}^{A}|}$ and normalizes the ratios to get the power generations’ change $\Delta P_{g}^{G}$. Here $ |\mathcal{G}^{A}|$ represents the cardinality of $ \mathcal{G}^{A}$. 
\begin{subequations}\label{eq:attack_sce}
	\begin{align}\label{eq:dpgg}
	\Delta P_{g}^{G} = \left[\sum_{l\in \mathcal{L}^A} \beta_{l}P_{l}^{L} \right] \times \frac{\lambda_g}{\sum_{g' \in \mathcal{G}^A} \lambda_{g'}} 
	\end{align}
All load changes $\Delta P_{l}^{L}$ and generation changes $\Delta P_{g}^{G}$, together with zeros that denote buses with unchanged injection make up the power injection change vector $\Delta P_{A}^{I} \in \mathbb{R}^{118}$. Besides, similar to \eqref{eq:z}, the attacker then utilizes the knowledge of the topology and grid parameters to coordinately calculate power flows change vector $\Delta P_{A}^{F} \in \mathbb{R}^{186}$. 
	\begin{align}\label{eq:dpaf}
	\Delta P_{A}^{F} = H^{F} \cdot \Delta P_{A}^{I}, 
	\end{align}
\end{subequations}
Afterwards, the attack vector $a$ consists of the change vector of loads, power generations and line power flows. 

The FDIA manipulates the original data of loads, power generations and line power flows. The pattern of the corrupted data may deviate from that of normal operating conditions, which enables it to be detected by the autoencoder if the reconstruction error $ {R}_{j}$ exceeds $\tau_{\alpha}$. 

\subsection{Load-targeted attack for economic profit} \label{subsec:scenario_1}

\subsubsection{Detection effectiveness validation}

\begin{figure}[t!p]
	\centering
	\includegraphics[scale=0.28]{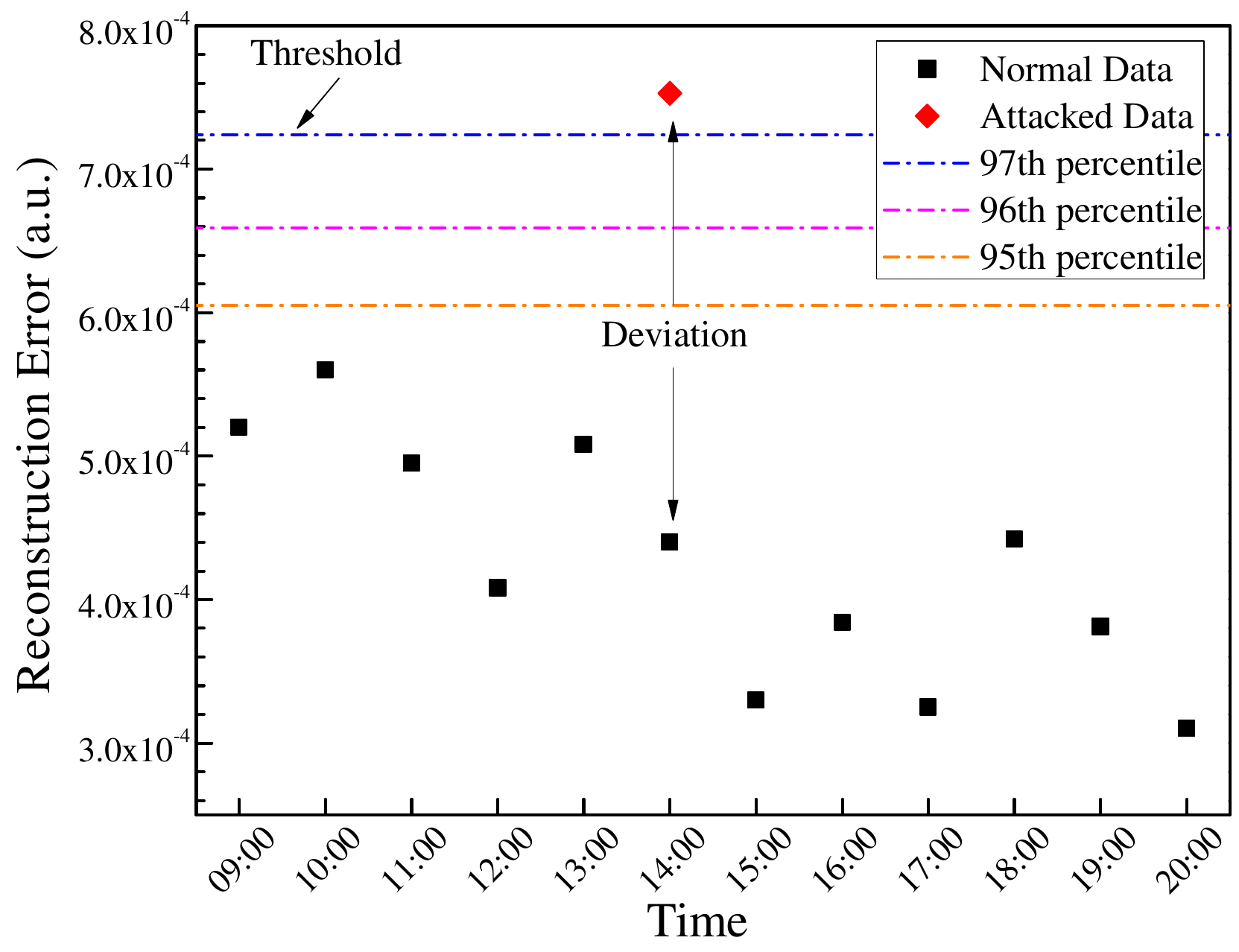}
	\caption{Detection effectiveness validation by launching an FDIA. 
	}
	\label{fig:example}
\end{figure}

We first validate the effectiveness of the trained detector. In this experiment, we observe the change of the reconstruction error $R_j$ before and after a false data injection attack and compare it with the threshold $\tau_{\alpha}$. A common scenario for an attack happens when the adversary gets the data of a local area and utilizes it to manipulate the neighboring measurements. Here, we select 12 hours’ operating data from 9:00 to 20:00 on December 31st, 2017 as an example. Assuming the attacker gets the three loads’ profile of bus 108, 109, 110, at 14:00, to gain economic profit, an attack is launched by injecting false data to decrease the power demand of the loads by 10$\%$ 
as $-7.48\, \mathrm{MW}$, $-5.69\, \mathrm{MW}$ and $-6.28\, \mathrm{MW}$ respectively. Accordingly, to balance the power of loads and generations, the attacker decreases the nearby power injection of two generators connected to bus number 110 and 111 with the ratio $\lambda_{1} : \lambda_{2} = 1$. Based on \eqref{eq:dpaf}, the corresponding transmission line power flows are obtained. The experiment result is depicted in Fig.~\ref{fig:example}.

From the result, we can observe that before the attack, the reconstruction error $R_j$ of normal operating data is in the range of $3.10 \times 10^{-4}$ and $5.60 \times 10^{-4}$, and they are lower than the threshold $\tau_{97\%} = 7.25\times 10^{-4}$ learned in the training process shown in the subsection B of Section~\ref{sec:detect}. To  be  specific, after  observing  the  reconstruction  error distribution  of  the validation  data,  the  threshold  is  set  as $97^{th}$ percentile  due  to the occurrence of the ‘inflection point’ where the cumulative distribution curve of the reconstruction error flattens out from the steep rise. After manipulation by the false data injection, the reconstruction error $R_j$ at 14:00 increases from $4.40 \times 10^{-4}$ to $7.53 \times 10^{-4}$, which exceeds the threshold $\tau_{97\%}$ and triggers an alarm. The detector thus recognizes an anomaly in the corrupted measurements, which deviate from measurements taken in normal operating conditions. This result demonstrates that the autoencoder is capable of FDIA detection in at least some scenarios. 

\subsubsection{General detection performance}

\begin{figure}[t!p]
	\centering
	\includegraphics[scale=0.28]{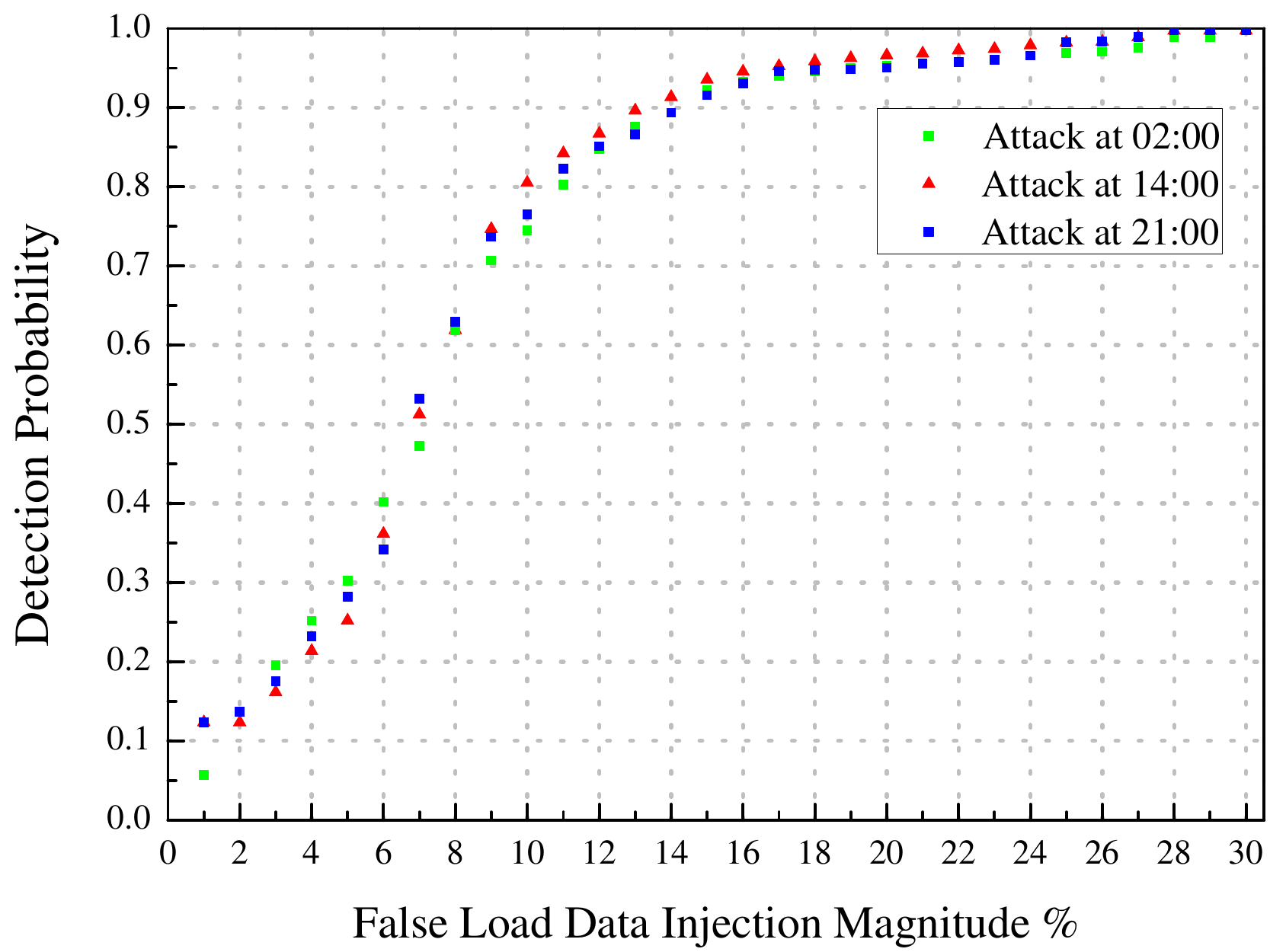}
	\caption{Detection probability of attacks at different time with different false load data injection magnitude.}
	\label{fig:differenttime}
\end{figure}

In addition to the one-off effectiveness demonstrated above, we are also interested in its statistical detection performance. This is tested by launching a larger number of FDIAs at various times and with various false load data injection magnitudes. Here the magnitude is defined as the percentage of load reduction in targeted nodes. For the sake of comparison, the attack targets remained the same as these utilized in the last experiment. In this experiment, we launch an attack at 2:00, 14:00 and 21:00 in each day of 2017 by reducing reported loads between 1$\%$ to 30$\%$ and observing the detection performance. The detection probability is the ratio of detected attacks to all the launched attacks, namely the true positive rate. The results are shown in Fig.~\ref{fig:differenttime}.

Because the load demands at 2:00, 14:00 and 21:00 differ significantly, the resulting power system states (including flows) are also substantially different.  However, the result shows, under the same false load injection magnitude, the detection probabilities differ only slightly. This demonstrates that the autoencoder learns the intrinsic relationship of the loads, power generations and power flows from different operating conditions, leading to robust detection results.

In addition, we launch 8760 attacks, one for each hour of 2017, by decreasing the power demand of the same buses by 15$\%$. Besides, we use the hourly normal operating data in 2017 as a control group. The result is shown in Table~\ref{tab:detection_prob}.

\begin{table}[h]
	\caption{Detection performance evaluation.}
	\centering
	\begin{tabular}{|p{0.10\textwidth}|p{0.095\textwidth}|p{0.10\textwidth}|p{0.095\textwidth}|}
		\hline
		& Normal Data &  & Attack Data\\
		\hline
		True Negative & $96.5\% \ (8453)$  & True Positive & $93.6\%$ \ (8199) \\
		\hline
		False Positive & $3.5\% \ (307)$  & False Negetive & $6.4\%$ \ (561)  \\
		\hline
	\end{tabular}
	\label{tab:detection_prob}
\end{table}

From the experiment result, we can find that the detection probability (true positive rate) is 93.6$\%$, which denotes a satisfactory detection performance. As mentioned in the first experiment, the threshold $\tau_{97\%}$ was used, corresponding to a 3\% misclassification rate in the validation set. It is worth noting that the false positive rate is comparable to the 3.5$\%$ observed in Table~\ref{tab:detection_prob}. This result suggests that the autoencoder has a good generalization capability and does not overfit.

\subsubsection{Detection performance comparison}

\begin{figure}[t!p]
	\centering
	\includegraphics[scale=0.055]{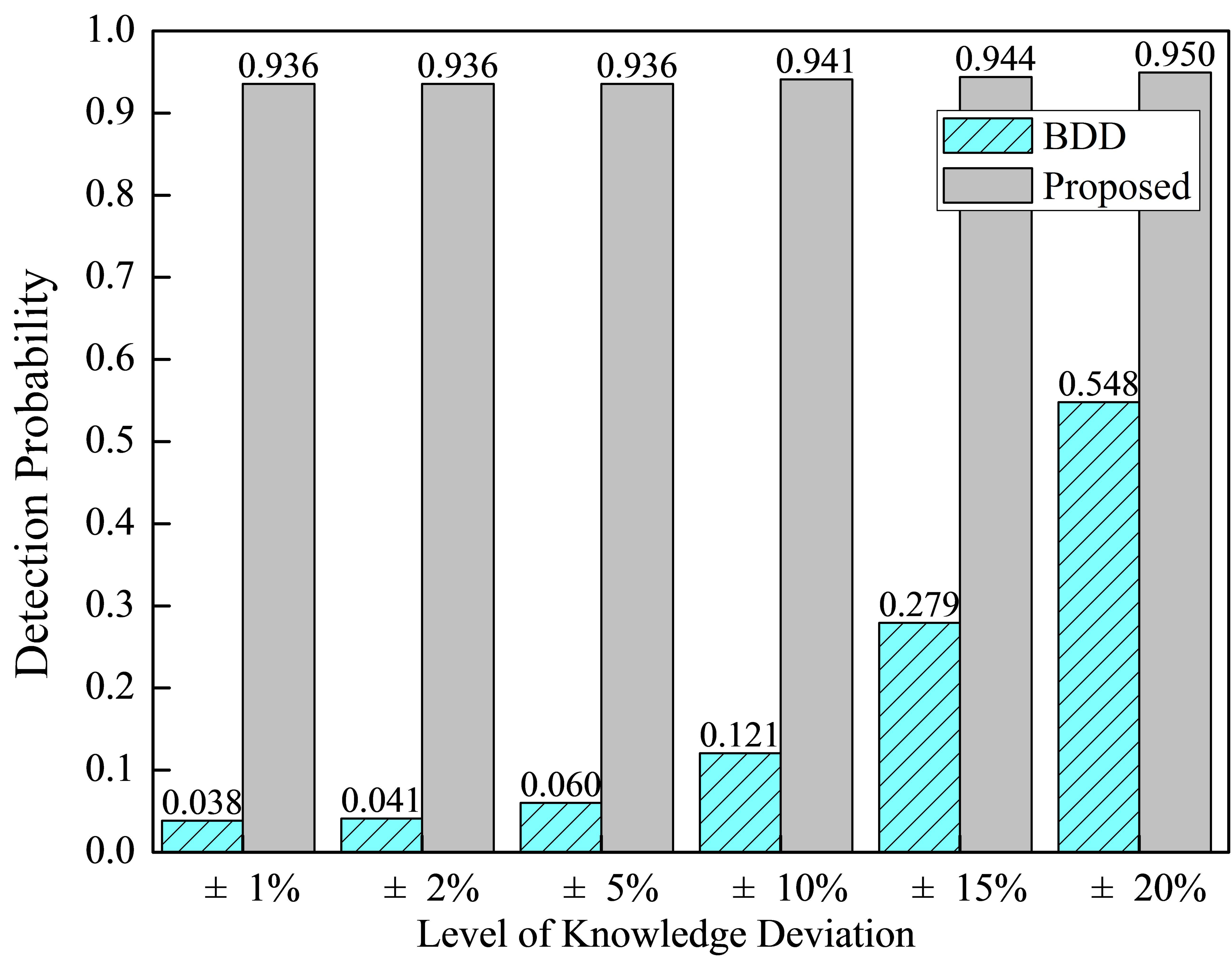}
	\caption{Detection performance comparison between the proposed mechanism and the BDD scheme in load-targeted attack scenario.}
	\label{fig:comparison}
\end{figure}

In the above experiments, our proposed autoencoder-based detector has succeeded in generating a diagnosis signal in the presence of FDIAs which can keep stealthy from the viewpoint of BDD. In the second experiment, we compare our detector with BDD in detection of `unstealthy' FDIAs. Such attacks have the possibility to be detected by the BDD while the detect ability is intimately related to the topology or parameter errors in the construction of FDIAs by the attacker. Thus in what follows there exist knowledge deviations in the system model acquired by the attacker in computing the attack vector of \eqref{eq:attack_sce}. In particular, we explore the case that the attacker knows the exact topology of the network but inaccurate line reactance $R$ in \eqref{eq:pipf}. This can be described by
\begin{align}\label{eq:hat_r}
\hat{R} = R \cdot (I^{R} + \gamma), 
\end{align}
where $I^{R} \in \mathbb{R}^{n_{t} \times n_{t}}$ is the identity matrix and $\gamma \in \mathbb{R}^{n_{t} \times n_{t}}$ is a diagonal matrix whose elements denote the reactance deviation ratio - which we will refer to as the \emph{knowledge deviation ratio}. In this experiment, we range the magnitude of the deviations from $0.01$ 
to $0.20$, with randomly sampled signs for each element. According to the explanation of \eqref{eq:z}, this will lead to an erroneous distribution factor matrix $H^{F}$ and thus obtain inaccurate power flow values. 
We keep the attack target unchanged from the previous experiments and set the false load data injection magnitude on the selected three loads by decreasing them by 15$\%$. The results are shown in Fig.~\ref{fig:comparison}. As the level of knowledge deviation increases from $\pm$1$\%$ to $\pm$20$\%$, the detection probability of BDD rises from 0.038 to 0.548, but it remains lower than the detection performance of the autoencoder. 

\section{Conclusion}
In this paper, we propose an FDIA detection mechanism based on an autoencoder neural network. The main contribution is that, distinct from existing approaches, the approach learns the internal dependency of ‘normal’ operation data, which avoids the need for gathering or generating attack data for training the classifiers and thus effectively overcomes the inherent unbalanced training data set challenge in power system. The results demonstrate that the mechanism is able to robustly detect stealthy FDIAs. Moreover, it still outperforms a BDD scheme when the attacker has only approximate knowledge of the network parameters. 

In future work, we aim to extend the method to analyze temporal signatures and to include contextual information.

\bibliographystyle{IEEEtran}
\bibliography{literature}

\end{document}